# Optimal prediction of decisions and model selection in social dilemmas using block models

Sergio Cobo-López[1], Antonia Godoy-Lorite[2], Jordi Duch[3], Marta Sales-Pardo[1*] and Roger Guimerà[1,4]

*Correspondence: marta.sales@urv.cat
[1]Departament d'Enginyeria Química, Universitat Rovira i Virgili, Tarragona, Spain
Full list of author information is available at the end of the article

**Abstract**

Advancing our understanding of human behavior hinges on the ability of theories to unveil the mechanisms underlying such behaviors. Measuring the ability of theories and models to predict unobserved behaviors provides a principled method to evaluate their merit and, thus, to help establish which mechanisms are most plausible. Here, we propose models and develop rigorous inference approaches to predict strategic decisions in dyadic social dilemmas. In particular, we use bipartite stochastic block models that incorporate information about the dilemmas faced by individuals. We show, combining these models with empirical data on strategic decisions in dyadic social dilemmas, that individual strategic decisions are to a large extent predictable, despite not being "rational." The analysis of these models also allows us to conclude that: (i) individuals do not perceive games according their game-theoretical structure; (ii) individuals make decisions using combinations of multiple simple strategies, which our approach reveals naturally.

## 1 Introduction

Many human activities have strong recurrent patterns that make them easy to predict [1, 2]. Other activities involve active decision making and are, therefore, not so obviously predictable. These include relatively simple decisions about, for example, which movie to watch or which book to purchase [3, 4], as well as complex strategic decisions in which individuals need to anticipate and take into consideration the decisions of others. For these complex strategic decisions, the question of whether the decision-making process is predictable, and to what extent, has been largely unexplored.

Strategic decision making has been studied in the social sciences, especially in political science and in economics [5, 6]. Experimental approaches, in which individuals play simplified games that pose specific social dilemmas, have been particularly insightful and have demonstrated that individuals often do not act "rationally" to maximize their profits [7–9]. This makes their behaviors more unpredictable than one may have anticipated. Unfortunately, approaches to analyze data from these experiments have focused mostly on characterizing aggregate behaviors (qualitatively or using regression-based approaches) and on measuring deviations from rationality, on the aggregate and at the level of individuals [10]. In general, however, they have not assessed quantitatively the power of existing theories to predict accurately the actions of each individual.

The lack of such analyses is significant because quantifying predictability provides a rigorous framework to compare theories of decision-making. Indeed, if we formalize theories into models and compare their predictive power, the most plausible theory will be, in general, the one that makes the most accurate predictions [11, 12]. Similarly, given a simple model that we aim to refine and improve, the refinements should increase predictability; otherwise, they ought to be revised or discarded [11]. All in all, studying the predictive power of theories and models opens the door to advance our understanding of human behavior on solid grounds [11].

Here, we aim precisely at narrowing this gap by proposing models for strategic decision making, by developing rigorous model inference approaches, and by showing that individual strategic decisions are, to a large extent, predictable. Specifically, we focus on a recent large-scale study of individuals playing a variety of dyadic games in a controlled setting [13]. We propose two models, and the corresponding inference approaches, that are more predictive than those built upon expectations of individuals' rationality. Our models are based on the assumption that there are groups of individuals that use similar decision-making strategies (have similar behavioral phenotypes [13–16]), and groups of games that are perceived similarly by individuals. We are agnostic a priori about which groups of individuals are most appropriate, so that the groups we obtain arise from fitting the models and are the ones that describe the observed behaviors most parsimoniously [12]. Similarly, we do not make strong assumptions about the groups of games that are perceived similarly by individuals and, in particular, we do not assume that games with the same Nash equilibrium [17] are in the same group. However, we do exploit existing information on the similarities between the payoffs of particular pairs of games. Importantly, our approach gives predictive models that are interpretable, which enables us to conclude that: (i) the perception of games by individuals is at odds with their game-theoretical structure; (ii) individuals do not use a single strategy when making decisions but rather a combination of multiple simple strategies, which our approach reveals naturally.

## 2 Data

We consider a dataset [13] consisting of 541 individuals playing a collection of dyadic games in which each player has to choose among two actions: cooperation ($C$) or defection ($D$). Each game is characterized by its payoff matrix (Fig. 1(a)): if both players cooperate, they both obtain a "reward" $R$; if both defect, they both get a "punishment" $P$; and if one cooperates and the other defects, the cooperator gets a "sucker's payoff" $S$ and the defector a "temptation payoff" $T$.

For the experiments, the reward and punishment were fixed to $R = 10$ and $P = 5$, whereas the other payoffs took values $S \in \{0, 1, \dots, 10\}$ and $T \in \{5, 6, \dots, 15\}$. Depending on these values, the games display different Nash equilibria [17] (which are the assumed rational behaviors in which no players have incentives to change behavior) and are typically divided into four groups (Fig. 1(b)) [13]: harmony game (HG), snowdrift game (SG), stag hunt (SH) game, and prisoner's dilemma (PD). HG and PD have a single stable equilibrium that corresponds to a pure strategy: cooperate (HG) and defect (PD). SH displays two equilibria

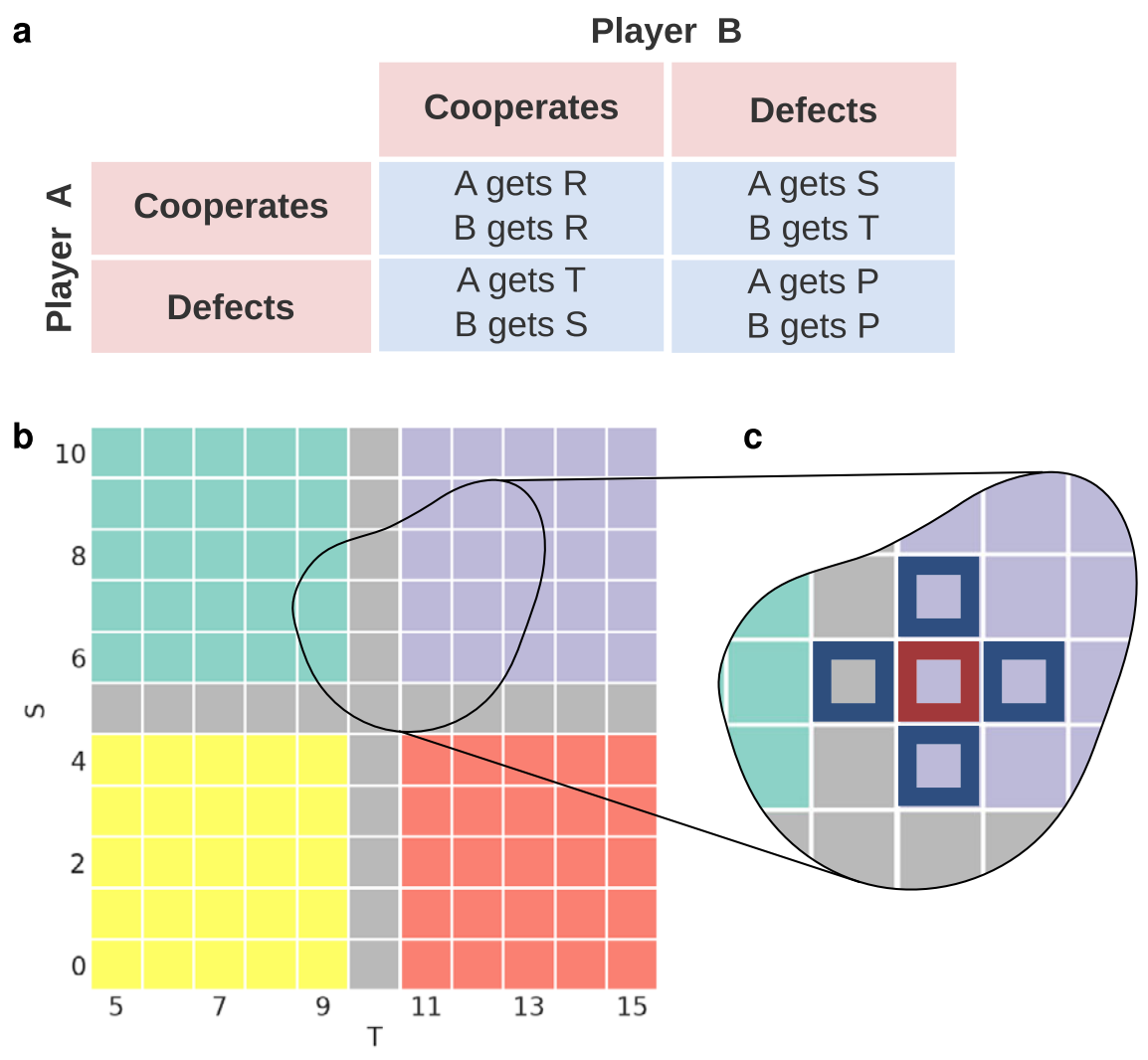

**Figure 1** Structure of the dyadic games. (**a**) Payoff matrix for the dyadic games played by individuals. (**b**) The games are typically divided into four groups depending on the different Nash equilibria and on the payoffs $S$ and $T$ (given that $R$ and $P$ are constant for all games). (**c**) In our models, we do not impose any group structure for games, but we assume that neighboring games are a priori more likely to belong to the same group. The neighbors of the game marked in red are marked in dark blue

corresponding to the two pure strategies. SG has a stable equilibrium that corresponds to a mixed strategy (that is, players have a certain probability of defecting and a certain probability of cooperating) [13, 17].

Each individual in the dataset played an average of 14 rounds, each one with a randomly selected payoff matrix (that is, a randomly chosen $(S, T)$ pair) and against a different, unknown, and randomly-selected player (that is, in general there were no repeated interactions between pairs of players). Based on the payoffs they obtained, participants received tickets for a lottery (one ticket for 40 payoff points), in which they could win four coupons of 50 euro redeemable at predefined stores [13].

We observe that the behavior of each player during the first four rounds is erratic, which leads to their behavior being less predictable during those rounds (Fig. S1). After round 4, all rounds are statistically indistinguishable by the metrics we use in what follows. Therefore, we discard the first four rounds of each player and consider all others as indistinguishable.

## 3 Single-strategy model, multiple-strategy model, and game metadata

It has been postulated that individuals can be classified into "behavioral phenotypes" depending on how they make decisions when facing social dilemmas [13–15]. The idea behind the concept of phenotype is that individuals apply general rules when making decisions, regardless of the structure of individual games. For example, players displaying the "envious" phenotype in Ref. [13] cooperate if, and only if, cooperation leads to higher payoffs for them than their opponents, regardless of whether that corresponds to the Nash equilibrium. In practice, among all possible combinations of strategies for single games there are only a few phenotypes that are followed by players. Here, we formalize this idea into group-based generative models of decision making, and go a step further to investigate to what extent these groups can be used to predict unobserved decisions.

We propose two models: a single-strategy model in which each player follows always the same strategy for the same game, and a multiple-strategy model in which players are not limited to following always the same strategy (Fig. S2). In the single-strategy model, we assume that each player $i$ belongs to one group of players $k$. Similarly, each game $g$ (defined by the payoff values $T$ and $S$) belongs to solely one group of games $\ell$. Moreover, the probability of player $i$ taking action $a_{ig} \in \{C, D\}$ in game $g$ depends exclusively on the groups $k$ and $\ell$, so that the probability of cooperation is

$$\Pr[a_{ig} = C] = p_{k\ell} \tag{1}$$

with $p_{k\ell} \in [0, 1]$. The element $p_{kl}$ is therefore the strategy of players in group $k$ for each of the games in group $\ell$. In a slight abuse of language we also call strategy the vector $\mathbf{p}_k$ of strategies for all game groups; because in the single-strategy model each individual has a single strategy, $\mathbf{p}_k$ also defines their phenotype. Note that in our approach we do not make any assumption about the strategies for individual games, which can either be pure ($p_{kl} \in \{0, 1\}$) or mixed ($0 < p_{kl} < 1$) [18].

In the multiple-strategy model, players do not always follow the same strategy but rather adopt different strategies with certain probabilities. We formalize this idea by allowing players to belong to a mixture of groups, with $\theta_{ik}$ being the probability that player $i$ belongs to group $k$ ($\sum_k \theta_{ik} = 1$), that is, that $i$ adopts strategy $k$. Similarly, we assume that games are not always regarded by players as belonging to the same group, and allow games to also belong to a mixture of groups; $\eta_{g\ell}$ is the probability that game $g$ is regarded as belonging to group $\ell$ ($\sum_\ell \eta_{g\ell} = 1$). In this model, the probability that player $i$ cooperates in game $g$ is

$$\Pr[a_{ig} = C] = \sum_{k,\ell} \theta_{ik} \eta_{g\ell} p_{k\ell}, \tag{2}$$

where $p_{k\ell}$ is the same as in the single-strategy model and the sum is over the $K$ groups for players and the $L$ groups for games. In fact, if we restrict the elements of the membership vectors $\boldsymbol{\theta}$ and $\boldsymbol{\eta}$ to be either 0 or 1 (that is games/players are restricted to belong to a single group) we recover the single-strategy model, so in what follows we use the multiple-strategy formulation without loss of generality.

A critical aspect in the modeling process (specially if we have limited data) is to specify how the *a priori* information we have about players and games affects the plausibilities of model parameters, namely the strategy matrices $\mathbf{p}$ and the group membership vectors $\boldsymbol{\theta}$ and $\boldsymbol{\eta}$. In our case, we only have auxiliary information (metadata) on the games. Therefore, we make no *a priori* assumptions about which values for $\mathbf{p}$ and $\boldsymbol{\theta}$ are more plausible (see Methods for details). In contrast, we expect games with similar payoffs $(S, T)$ to be regarded by players as similar. This means that games that are neighbors in the $TS$-plane are more likely to have similar membership vectors. We model this expectation by choosing a prior distribution that introduces an exponential penalty when membership vectors of neighboring games are dissimilar

$$P(\boldsymbol{\eta}) \propto \exp\left[-\alpha \sum_{\langle gg' \rangle} (1 - \eta_g \cdot \eta_{g'})\right]. \tag{3}$$

Here, the sum runs over all pairs of games that are nearest-neighbors in the $TS$-plane, that is, that differ by plus or minus one in $S$ or $T$ (but not both; Fig. 1(c)), and $\alpha \geq 0$ is a parameter that we call the game aggregation factor. Note that $\alpha$ plays a similar role as the interaction in Ising, Potts and N-spin models [19, 20], so that for $\alpha = 0$ we recover the uniform prior, whereas increasing values of $\alpha$ make it more likely that neighboring games have similar mixing vectors. Note that in the single-strategy model, Eq. (3) is in reality a prior over partitions of games into groups $\pi_g$, since, as for the players, the model considers only the set of membership vectors $\boldsymbol{\eta}$ that result in disjoint partitions.

These models are reminiscent of existing models, but differ in important aspects. The single-strategy model is the formalization of the idea that there exist "behavioral phenotypes," and that decisions depend only on those phenotypes [13–15]. Unlike previous work, however, we do not assume that the groups of games are known *a priori*. Formally, this model is a bipartite stochastic block model [21], whereas the multiple-strategy model is a bipartite mixed-membership stochastic block model [4, 22, 23]. The idea of allowing the existence of more than one strategy is reminiscent of the population strategy models used in evolutionary game theory in which competing strategies can coexist within a population [18]. The difference is that, in our case, each individual is allowed to simultaneously consider more than one strategy to make decisions, rather than having a population in which different strategies are represented. Moreover, as we previously mentioned, in our approach we make no assumptions about which strategy (pure or mixed) players are using in each game; we can determine whether players are using mixed strategies or not as an outcome of the inference process. This possibility is especially interesting in the analysis of real data since the empirical relevance of this concept has been so far hard to prove [24, 25].

Also unlike previous models, we are able to use the available game metadata, which we introduce in the model through the prior distribution of model parameters, an approach that is reminiscent of what has been used in networks [26] (an alternative is to model the metadata together with the data [27]). Modeling game similarity through the scalar product of game membership vectors opens the door to modeling other network-like systems whose nodes, like games, are embedded in a low-dimensional space.

## 4 Inference of the most plausible partitions and mixtures

Given a record of observed actions $A^o$ from many players, our goal is to find the most plausible membership vectors (or, in the case of the single-strategy game, to find the most plausible partition into groups of players and games). The most plausible membership vectors are those that maximize the posterior $P(\boldsymbol{\theta}, \boldsymbol{\eta}|A^o)$ and, in general, they are also the most predictive ones [12]. Next, we show how to obtain them in both models.

For the single-strategy model, it is possible to analytically marginalize the posterior over the strategy matrices $\mathbf{p}$, which gives (Additional file 1, Sect. 3)

$$P(\boldsymbol{\theta}, \boldsymbol{\eta}|A^0) = \frac{1}{Z}\exp\bigl[-\mathcal{H}(\boldsymbol{\theta}, \boldsymbol{\eta})\bigr]. \tag{4}$$

Here, $Z$ is a normalizing constant, and $\mathcal{H}(\boldsymbol{\theta}, \boldsymbol{\eta})$ is given by

$$\mathcal{H}(\boldsymbol{\theta}, \boldsymbol{\eta}) = \sum_{k\ell}\bigl[\ln(n_{k\ell}+1)! - \ln\bigl(n^C_{k\ell}\bigr)! - \ln\bigl(n^D_{k\ell}\bigr)!\bigr] + \alpha F \tag{5}$$

with $n^C_{k\ell}$ and $n^D_{k\ell}$ being the number of observed cooperations and defections, respectively, of players in group $k$ in games in group $\ell$, $n_{k\ell} = n^C_{k\ell} + n^D_{k\ell}$ the number of observed actions for those groups, and $F$ the number of neighboring pairs of games that are not in the same group. We obtain the most plausible partitions $(\theta^*, \eta^*)$ of players and games by minimizing $\mathcal{H}$, that is

$$(\theta^*, \eta^*) = \arg\min_{\theta,\eta} \mathcal{H}(\boldsymbol{\theta}, \boldsymbol{\eta}). \tag{6}$$

Because this is a combinatorial optimization problem, we use simulated annealing for the minimization (Additional file 1, Sect. 4). Note that we do not have to fix the number of groups for players and games in the single-strategy model; we obtain these groups as a result of the inference process.

For the multiple-strategy model, we cannot marginalize over the strategy matrices exactly. Therefore, we assume that the distribution $P(\boldsymbol{\theta}, \boldsymbol{\eta}, \mathbf{p}|A^0)$ is very peaked around $\mathbf{p}^* = \arg\max_{\mathbf{p}} P(\boldsymbol{\theta}, \boldsymbol{\eta}, \mathbf{p}|A^0)$ and use the approximation [4]

$$P(\boldsymbol{\theta}, \boldsymbol{\eta}|A^0) \approx P(\boldsymbol{\theta}, \boldsymbol{\eta}, \mathbf{p}^*|A^0). \tag{7}$$

Because $\boldsymbol{\theta}$ and $\boldsymbol{\eta}$ are not discrete as in the single-strategy model, one can use a variational approach to obtain analytic expressions for the optimal membership vectors $\theta^*$ and $\eta^*$, as well as for $\mathbf{p}^*$.

We obtain the following equations for the model parameters (Additional file 1, Sect. 3)

$$\theta_{ik} = \frac{\sum_{g\in\mathcal{C}_i}\sum_{\ell} w_{ig}^C(k,\ell) + \sum_{g\in\mathcal{D}_i}\sum_{\ell} w_{ig}^D(k,\ell)}{d_i}, \tag{8}$$

$$\eta_{g\ell} = \frac{\sum_{i\in\mathcal{C}_g}\sum_{k} w_{ig}^C(k,\ell) + \sum_{i\in\mathcal{D}_g}\sum_{k} w_{ig}^D(k,\ell)}{d_g + \alpha\sum_{r\in\partial g}\eta_r\cdot\eta_g} + \frac{\alpha\sum_{r\in\partial g}\eta_{r\ell}\eta_{g\ell}}{d_g + \alpha\sum_{r\in\partial g}\eta_r\cdot\eta_g}, \tag{9}$$

$$p_{k\ell} = \frac{\sum_{(ig)\in\mathcal{C}} w_{ig}^C(k,\ell)}{\sum_{(ig)\in\mathcal{C}} w_{ig}^C(k,\ell) + \sum_{(ig)\in\mathcal{D}} w_{ig}^D(k,\ell)}. \tag{10}$$

Here $\mathcal{C}$ is the set of observed player-game pairs such that $a_{ig} = C$, and $\mathcal{D}$ is the set of observed player-game pairs such that $a_{ig} = D$. Correspondingly, $\mathcal{C}_i/\mathcal{D}_i$ is the set of observed games in which player $i$ cooperates/defects, and $\mathcal{C}_g/\mathcal{D}_g$ is the set of players that cooperate/defect in game $g$. $d_i$ is the total number of games played by player $i$, $d_g$ is the total number of players that played game $g$, and $\partial g$ is the set of the nearest-neighbors of game $g$ in the $TS$-plane. Finally, $w_{ig}^C(k\ell)$ and $w_{ig}^D(k\ell)$ are the estimated probabilities that a specific action (cooperate or defect) is due to player $i$ and game $g$ belonging to groups $k$ and $\ell$ respectively, which can be computed as

$$\begin{aligned} w_{ig}^C(k,\ell) &= \frac{\theta_{ik}\eta_{g\ell}p_{k\ell}}{\sum_{k'\ell'}\theta_{ik'}\eta_{g\ell'}p_{k'\ell'}}, \\ w_{ig}^D(k,\ell) &= \frac{\theta_{ik}\eta_{g\ell}(1-p_{k\ell})}{\sum_{k'\ell'}\theta_{ik'}\eta_{g\ell'}(1-p_{k'\ell'})}. \end{aligned} \tag{11}$$

The equations can be solved iteratively using an expectation-maximization (EM) algorithm in order to find the optimal model parameter values [4]. Because iterating the EM algorithm of the update equations can lead to different fixed points depending on its initial conditions, we perform 500 independent runs and take the run with the maximum a posteriori to make predictions. In contrast to the single-strategy model, here we have to fix $K$ and $L$ and use model selection criteria to select the optimal values (Additional file 1, Sect. 5 and Fig. S3).

## 5 Baseline model

We start by studying the predictive power of the model proposed in Ref. [13], which we will consider as a baseline prediction. In this model, games are divided into four fixed groups (harmony game, snowdrift game, stag hunt game, and prisoner's dilemma). Each user $i$ is then characterized by a strategy vector $v_i$ that quantifies their propensity to cooperate in each of the four types of games. In the baseline model, players are grouped according to the similarity in their strategy vectors using $k$-means.

To test the predictive power of the model, we use 5-fold cross-validation, that is, we divide the data in five equally-sized splits, and then use four splits as a training set and the remaining split as a test set to assess the capacity of the model to predict unobserved data. We repeat this for the five possible train-test combinations. For each training set, we find the player groups, and estimate $p_{k\ell}$, the probability that players in group $k$ cooperate in games in group $\ell$, as the frequency with which players in group $k$ cooperate in games in group $\ell$ in the empirical data. Then, we use these frequencies to predict cooperation in the test data, so that if $p_{kl} > 0.5$ then the prediction is that all users in group $k$ will cooperate in games in group $\ell$. We obtain an average predictive accuracy of $0.683 \pm 0.005$ (Fig. 2(a)).

## 6 Single-strategy models: maximally predictive partitions reveal perception of games by players

Next, we study the predictive power of the single-strategy model as a function of the game aggregation parameter $\alpha$, which controls how strongly neighboring games are pushed into the same group. We use the same 5-fold cross validation scheme as before. For each split, we obtain the optimal partitions of players and groups from Eq. (6) and, as before, use the observed cooperation frequencies of groups of players in groups of games to make predictions on the test set.

We find that the predictive power of the model increases with $\alpha$ and reaches its maximum for $\alpha = 2$, at which point it is significantly more predictive than the baseline model with a predictive accuracy of $0.714 \pm 0.008$ (Fig. 2(a)).

A close inspection of the optimal partitions for players and games reveals that the game aggregation factor has an effect on the partition of both players and games into groups (Fig. 2(b), (c)–(f)). With respect to the number of groups of players, we observe that the number of groups decreases as we increase $\alpha$, and stabilizes for $\alpha > 2$ at around 20 groups. Note that many of these player groups are small since 5 or 6 groups typically account over 50% of the players (see bottom row of Fig. 2(c)–(f)). With respect to the partitions of games we observe two noteworthy aspects. First, that the absence of a prior for game memberships ($\alpha = 0$) makes game groupings (and as a surrogate player's groupings) too susceptible to fluctuations, which results in low predictive power. Second, that as $\alpha$ increases the prior helps disregard statistical fluctuations in favor of a well-defined structure of game groups within the $TS$-plane, leading to a higher predictive performance. Despite the fact that the prior acts on the games alone, it also leads to a lower number of player groups because with fewer game groups the number of possible strategy vectors is also smaller. Interestingly, games fall into groups defined by the difference between sucker and temptation payoffs $\Delta = (T - S)$, and are qualitatively different from the four regions that follow from game-theoretic considerations (Fig. 1(b)). More specifically, at the optimal aggregation factor $\alpha = 2$ we observe three regions that correspond approximately to: (i) $S > T$, where most players cooperate; (ii) $S < T - P = T - 5$, where most players do not cooperate (although some do, including a few that always cooperate); (iii) the intermediate region where some cooperate and others do not. These groups are consistent with the observations described in Ref. [13], but in our analysis arise naturally from the rigorous comparison of models, and get incorporated in the models. In this sense, our approach illuminates the way in which individuals are "predictably irrational" [28].

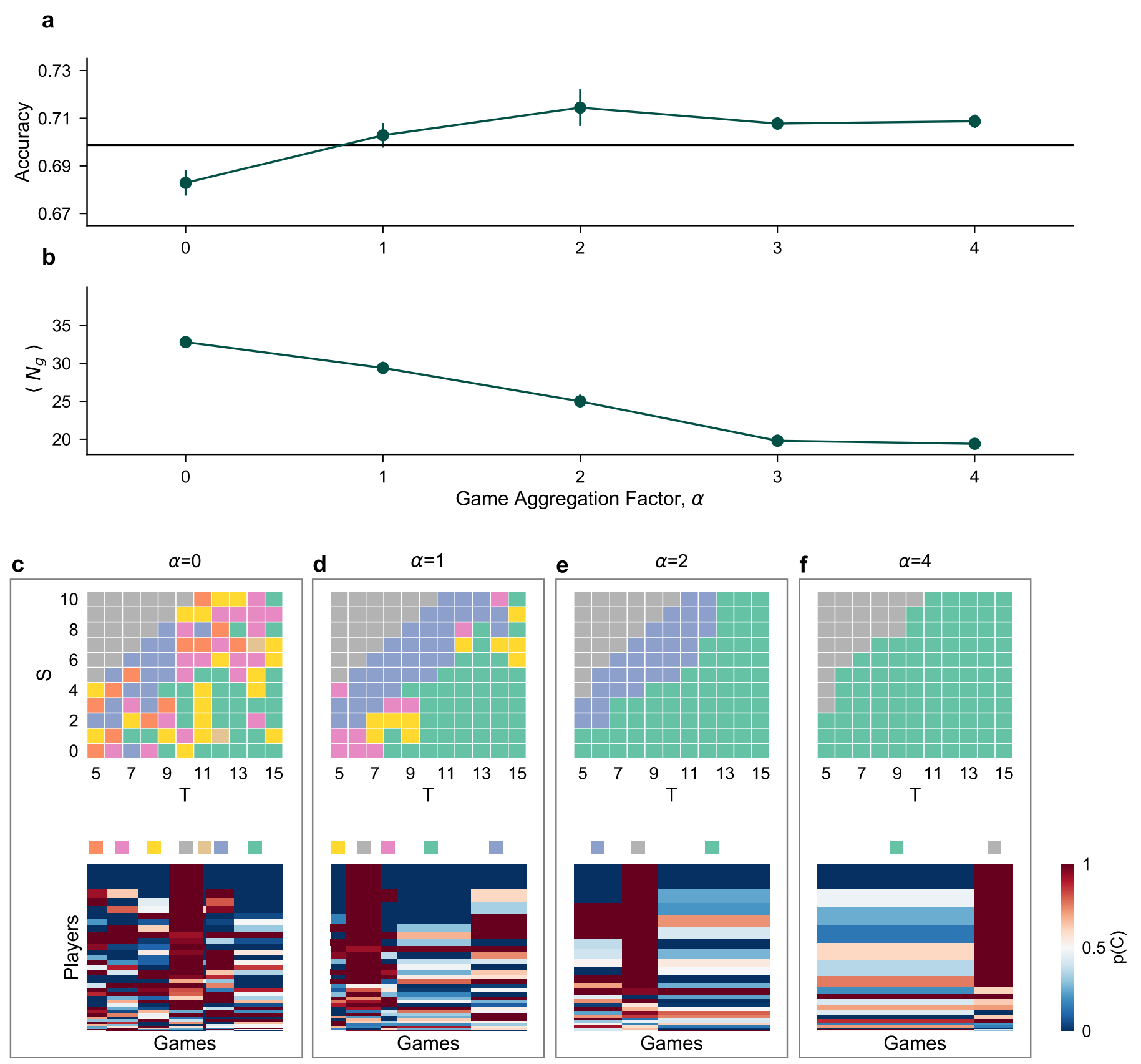

**Figure 2** Single-strategy model. (**a**) Predictive accuracy of the single-strategy model as a function of the game aggregation factor $\alpha$. Each point represents the average of a 5-fold cross-validation; error bars indicate the standard error of the mean. The solid black line represents the accuracy of the baseline model (see text and Ref. [13]). (**b**) Average number of groups of players $\langle N_g \rangle$ as a function of the game aggregation factor $\alpha$. Each point represents the average of the number of groups for players identified for each of the 5 folds; error bars indicate the standard error of the mean. (**c**)–(**e**) Groups of games in the *TS*-plane (top) and cooperation matrix **p** (bottom) for aggregation factors $\alpha$ equal to: (**c**) 0, (**d**) 1, (**e**) 2, (**f**) 4. In the game plots, each color indicates a different group of games in the most plausible partition as obtained from Eq. (6) for each value of $\alpha$. In the cooperation matrices, each row corresponds to a group of players $k$ (with height proportional to the number of players in the group), and each column to a group of games $\ell$ as indicated by the colors at the top (with width proportional to the number of games in the group). Each element $p_{k\ell}$ represents the probability that an individual in group $k$ cooperates when playing a game in group $\ell$, with dark red meaning always cooperate and dark blue always defect. The game groups and the cooperation matrices correspond to one of the folds in the 5-fold cross-validation, but are very consistent across folds (see Fig. S5)

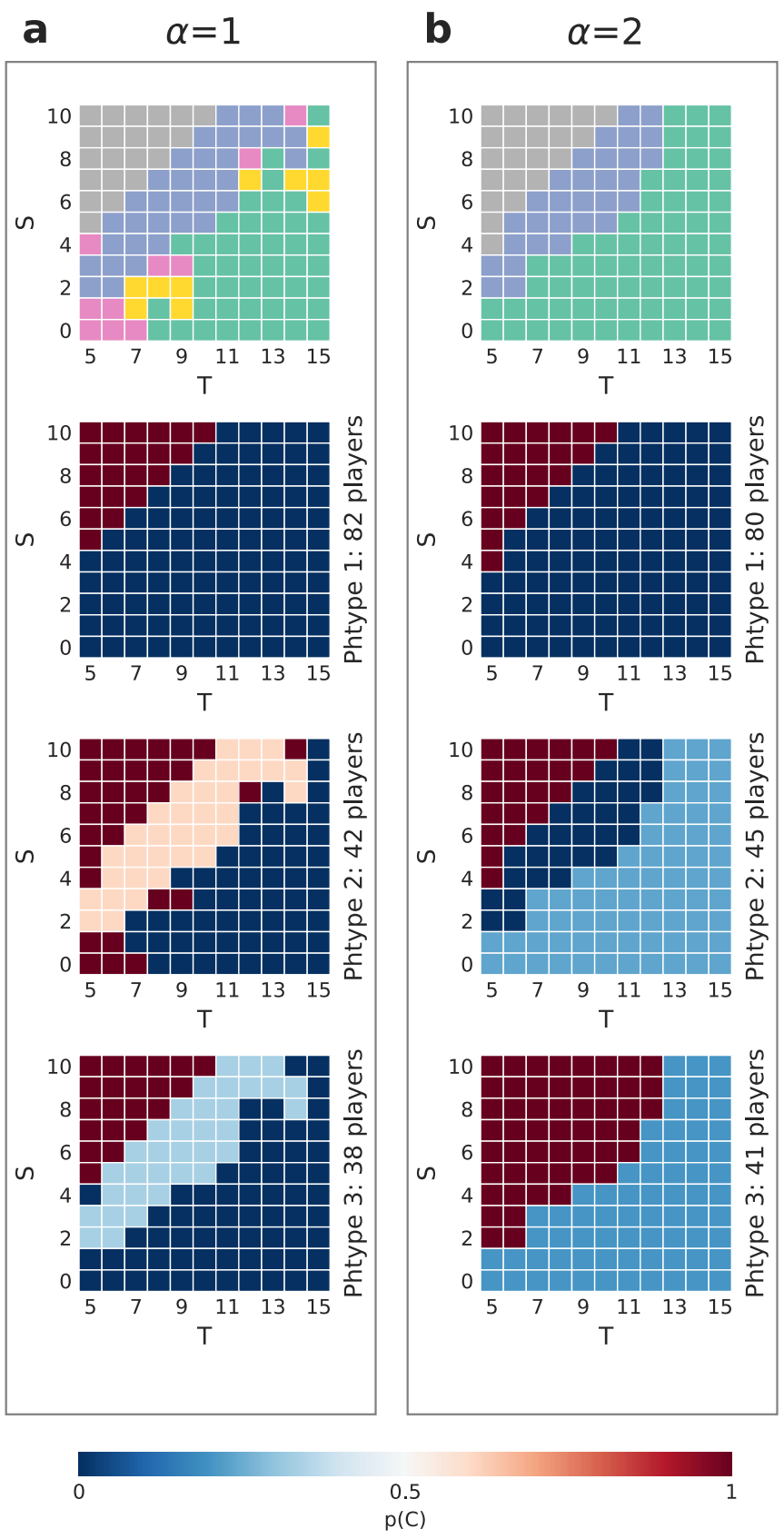

**Figure 3** Most common behavioral phenotypes for single-strategy models. Game groups (top; as in Fig. 2), and cooperation probability in each game group for the three largest player groups (or behavioral phenotypes) for: (**a**) $\alpha = 1$, (**b**) $\alpha = 2$. The groups and cooperation probabilities correspond to one of the folds in the 5-fold cross-validation, but are very consistent across folds

To further investigate the collective behavior of players, we focus on the phenotypes associated to the three largest groups of players identified for $\alpha = 1, 2$ (Fig. 3). In both cases we see that the largest group is characterized by two facts: i) players only distinguish between two types of games ($S \geq T$ and $S < T$); and ii) players display pure strategies in these games: always cooperate in games with $S \geq T$ and always defect in games with $S < T$ (Fig. 3). Interestingly, this phenotype is precisely the envious phenotype identified in [13], but in our case it arises naturally from our model-selection criteria without having to make any assumptions about the structure of the $TS$-plane or about the number of player groups. The two remaining most common phenotypes for $\alpha = 1, 2$ also show that players fully cooperate in games with $S \geq T$. However, the cooperation patterns for $S < T$ cannot be mapped directly into any behavioral phenotype described previously in the literature, and, despite being strongly correlated with $\Delta$ and leading to better predictions, they are not as easy to interpret as the most common phenotype. In part, this is due to the large number of observed phenotypes (Fig. 2(c)–(f)); in the following section, we show that the multiple-membership model provides a more parsimonious and straightforward description of this variety of phenotypes.

## 7 Multiple-strategy models are more predictive and easier to interpret than single-strategy models

Finally, we investigate the predictive power of models in which players are allowed to use mixtures of strategy vectors. As we show in Fig. 4, we find that, again, predictive

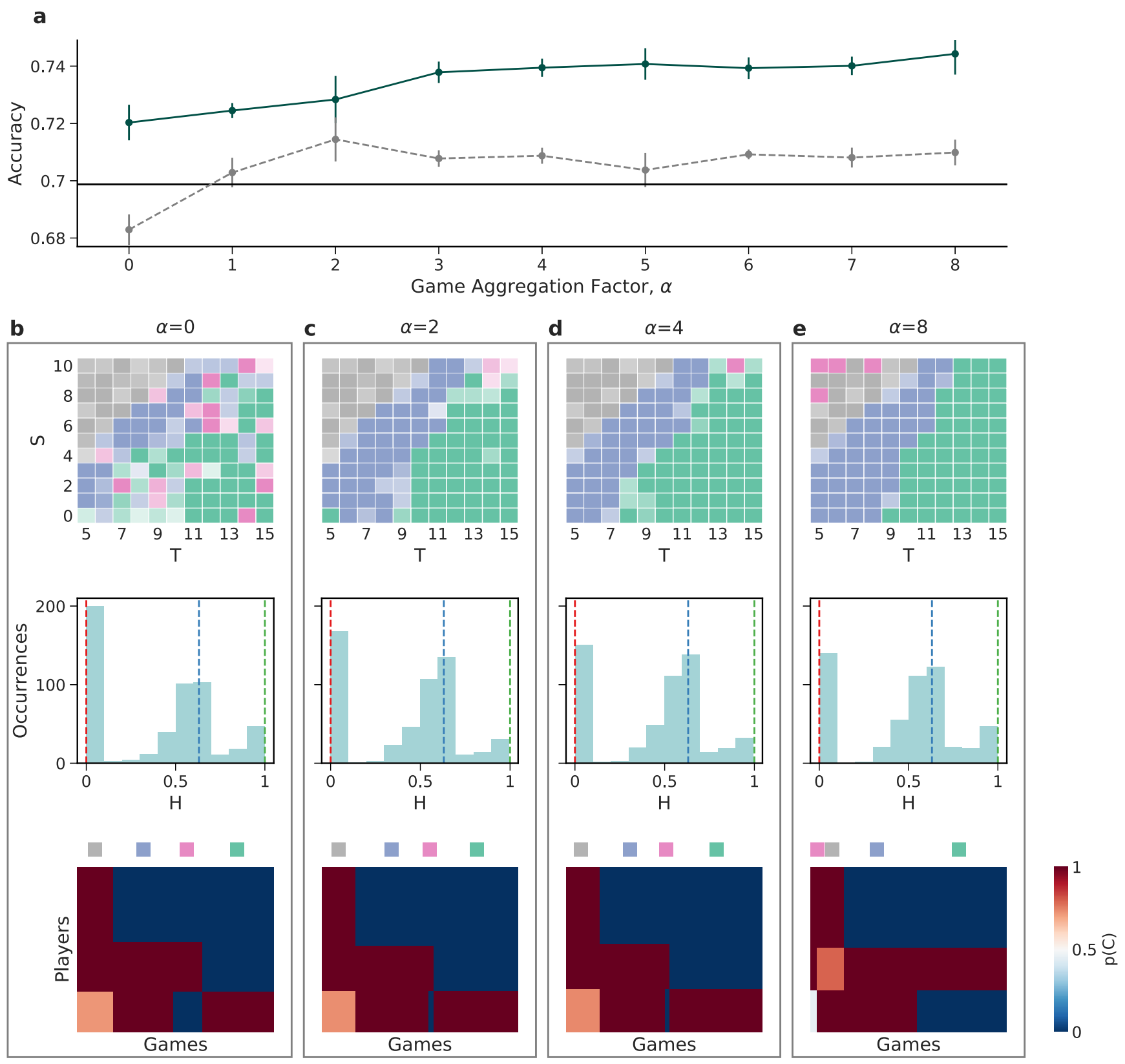

**Figure 4** Mixed-strategy model. (**a**) Predictive accuracy of the multiple-strategy model (green) as a function of the game aggregation factor $\boldsymbol{\alpha}$. We show results for $K = 3$ latent groups of players and $L = 4$ latent groups of games; increasing the number of groups did not result in an increase in accuracy (see Fig. S3). Each point represents the average of a 5-fold cross-validation; error bars indicate the standard error of the mean. Grey symbols represent the predictive accuracy of the single-strategy model as in Fig. 2(a), and the solid black line represents the accuracy of the baseline model. (**b**)–(**e**) Top group memberships for each game (top), distribution of the entropies of player membership vectors (middle), and cooperation matrices **p** (bottom) for each of the values $\boldsymbol{\alpha} = 0, 2, 4, 8$. In the game membership plots, each color indicates a different group of games in the most plausible model as obtained from Eqs. (8)–(11). The saturation of the color indicates how distributed a game is among groups, so that games with multiple memberships (typically at the border between groups) are paler (Fig. S4). In the distribution of the player entropies, players with a single strategy vector entropy $H = 0$ (red dashed line); players that mix two strategy vectors with equal weights have entropy $H = \log_3 2 = 0.63$ (blue dashed line); and players that mix three strategy vectors with equal weights have entropy $H = 1$ (green dashed line). In the cooperation matrices **p**, each row corresponds to a group of players $k$, and each column to a group of games $\ell$ as indicated by the colors at the top. The height of player group $k$ is proportional to the effective number of players in that group $\sum_i \theta_{ik}$. The width of game group $\ell$ is proportional to the effective number of games in that group $\sum_g \eta_{g\ell}$. Each element $p_{k\ell}$ represents the probability that an individual in group $k$ cooperates when playing a game in group $\ell$, with dark red meaning always cooperate and dark blue always defect. The groups, entropies, and cooperation matrices correspond to one of the folds in the 5-fold cross-validation, but are very consistent across folds (see Fig. S6)

performance grows with the game aggregation $\alpha$ and saturates after $\alpha = 3$. Remarkably, the multiple-strategy model is, in all cases, significantly more predictive than the single-strategy model, with a maximum predictive accuracy of $0.744 \pm 0.007$.

Let us consider first the effect of mixed-membership on game grouping. We observe that $\alpha$ has a smaller impact in the group membership vectors of games than it has in single-strategy models (Fig. 4(b)). Indeed, for all values $\alpha > 0$ the top membership matrix for games strongly resembles the game classification for single-strategy models in Fig. 2, which confirms that the perception of games by players following the difference $\Delta = (T - S)$ is robust. Rather than changing the structure of game groups, the main effect of increasing $\alpha$ is increasing the localization of games into groups—the top group membership of most of the games tends to one, whereas the other memberships all vanish to zero (see also Fig. S4). This implies that mixed-membership does not play a major role in game grouping, since games end up belonging mostly to a single well-defined group as in the single-strategy model (Fig. 4(b)).

Therefore, the increase in predictive performance must be due to the multiple membership of players, that is, to the fact that players are best described as not making decisions following a unique strategy but rather using a combination of strategies. Indeed, we find that the majority of players use a mixture of strategies (center row in Fig. 3), and that three global strategies are enough to make the most accurate predictions of players' decisions (bottom row in Fig. 3). Note that, unlike the single-strategy model, we need to fix the number of groups; but we find that such a small number of strategies provides the most accurate predictions (Fig. S3). Interestingly, these global strategy vectors are for the most part combinations of pure strategies in which players either fully cooperate or fully defect in games.

All three global strategies are used often (Fig. 4(b)–(e)), although one of them, the envious strategy (to cooperate only in games with $S \geq T$), is slightly most common than the others. This, again, is consistent with the results in Ref. [13] and with the most common strategy obtained using the single-strategy model. The other two strategies correspond to: (i) a more rational strategy that leads to cooperation for half of the games (including all harmony games) and to defection for the other half (including all prisoner's dilemma games); (ii) a strategy that accounts for non-rational behaviors, including incomplete cooperation in harmony games and full cooperation for most other games, including prisoner's dilemma games. This last strategy may seem counterintuitive but arises from the need to assign non-zero probability to all behaviors; without this strategy, for example, any observed non-cooperation in games with $S \geq T$ would lead to zero likelihood. It may seem surprising, however, that this deviant strategy is used in as many as 25% of all the decisions.

## 8 Discussion and conclusions

We have explored the power of group-based models to predict decisions made by individuals in simple classes of dyadic games that involve strategic thinking. Such decisions are known to deviate from the rationally expected behavior. However, our analysis proves that they still are highly predictable (74% of the decisions can be correctly predicted) and that group-based models are good models of strategic decision making.

More importantly, proposing interpretable models of human behavior and comparing them in terms of their predictive accuracy sets the bases for advancing the social sciences on solid grounds [11]. In this regard, we have shown that the most explanatory groupings of games reveal the perception of games by players, which differs from game-theoretical expectations. Our approach also gives the most explanatory cooperation strategies followed by players, and suggests that models in which players are allowed to use multiple strategies (rather than sticking to a single strategy) are more predictive than those models in which players are restricted to a single strategy. Multiple-strategy models are also more parsimonious in that they summarize the wide variety of phenotypes suggested by single strategy models as combinations of a small number of simple strategies. In fact, the combination of these two factors (perception of games by players and multiple strategies) accounts well for the rich variety of phenotypes observed in real data.

More broadly, we believe that our approach and models can be used to analyze many other behavioral experiments and datasets. Indeed, whenever humans face distinct (discrete, non-overlapping) situations and take distinct actions, their decisions can be modeled using the exact same approaches we have proposed here. For example, one could model how individuals make decisions in stock markets based on the situation they face (for example market going up or down). Given the expressiveness of group-based models, we anticipate that such models would provide accurate predictions and insightful characterization of behaviors. All together, we think that our work could have important implications in how new experiments, models, and theories are built to better understand human decision making.

## Additional material

**Additional file 1:** Supplementary materials (PDF 3.9 MB)

**Acknowledgements**
We thank J. Poncela-Casasnovas and A. Sánchez for helpful comments and suggestions.

**Funding**
This work was supported by the Ministerio de Economia y Competitividad de España (FIS2016-78904-C3-1-P).

**Abbreviations**
C, Cooperation; D, Defection; S, Sucker's payoff; T, Temptation payoff; HG, Harmony game; SG, Snowdrift game; SH, Stag hunt game; PD, Prisoner's dilemma; EM, Expectation-maximization.

**Availability of data and materials**
The data that support the results of this project are available in the Zenodo Repository (https://doi.org/10.5281/zenodo.1127154).

**Competing interests**
The authors declare that they have no competing interests.

**Authors' contributions**
SC-L, AG-L, JD, RG, and MS-P designed the research; SC-L and AG-L wrote the software for the data analysis; SC-L, AG-L, JD, RG, and MS-P discussed the results of the data anlysis; SC-L, RG, and MS-P wrote the manuscript. All authors read and approved the final manuscript.

**Author details**
[1]Departament d'Enginyeria Química, Universitat Rovira i Virgili, Tarragona, Spain. [2]Department of Mathematics, Imperial College London, London, United Kingdom. [3]Departament d'Enginyeria Informàtica i Matemàtiques, Universitat Rovira i Virgili, Tarragona, Spain. [4]ICREA, Barcelona, Spain.

## Publisher's Note